\documentclass[english,aps,prstper,reprint,showpacs,longbibliography,superscriptaddress]{revtex4-2}   

\usepackage[T1]{fontenc}	
\usepackage[latin9]{inputenc}	
\usepackage{geometry}		
\usepackage{graphicx}
\usepackage{times}

\usepackage{hyperref}  
\hypersetup{colorlinks=true,urlcolor=blue,citecolor=blue,linkcolor=blue}   
\usepackage{enumitem}          
\setlist{nosep}                 

\usepackage[capitalise,noabbrev]{cleveref} 

\usepackage{multirow}

\usepackage{caption}
\usepackage{subcaption}

\begin{document}

\title{More or better data: A new perspective on student reasoning about measurement uncertainty}

\author{Andy Schang}
  \affiliation{Laboratory of Atomic and Solid State Physics, Cornell University, 245 East Avenue, Ithaca, NY,
    14853} 
  \author{Matthew Dew}
  \affiliation{Laboratory of Atomic and Solid State Physics, Cornell University, 245 East Avenue, Ithaca, NY,
    14853} 
  \author{Emily M. Stump}
  \affiliation{Laboratory of Atomic and Solid State Physics, Cornell University, 245 East Avenue, Ithaca, NY,
    14853} 
\author{N. G. Holmes}
   \affiliation{Laboratory of Atomic and Solid State Physics, Cornell University, 245 East Avenue, Ithaca, NY,
    14853} 
  \author{Gina Passante}
  \affiliation{Department of Physics, California State University Fullerton, 800 N. State College Blvd., Fullerton, CA,
    92831} 

\begin{abstract}

Uncertainty is an important and fundamental concept in physics education. Students are often first exposed to uncertainty in introductory labs, expand their knowledge across lab courses, and then are introduced to quantum mechanical uncertainty in upper-division courses. This study is part of a larger project evaluating student thinking about uncertainty across these contexts. In this research, we investigate advanced physics student thinking about uncertainty by asking them conceptual questions about how a hypothetical distribution of measurements would change if `more' or `better' data were collected in four different experimental scenarios. The scenarios include both classical and quantum experiments, as well as experiments that theoretically result in an expected single value or an expected distribution. This investigation is motivated by our goal of finding insights into students' potential point- and set-like thinking about uncertainty and of shining light on the limitations of those binary paradigms.

\end{abstract}
\maketitle

\section{Introduction}

The concept of uncertainty is a fundamental aspect of physics~\cite{heron_phys21_2016}, particularly in undergraduate instructional laboratories (labs)~\cite{american_association_of_physics_teachers_aapt_2014}. While many physics lab instructors cite uncertainty-related goals for their courses, the format of these goals range from procedural (e.g., carrying out procedures to propagate uncertainties or reporting measurements with uncertainties), conceptual (e.g., describing how standard deviation captures the variability between trials or rounding limits inform uncertainty in single measurements), to more agentic (e.g., deciding what are the major sources of uncertainty in an experiment and designing experiments to minimize those sources). The physics education research community has also captured students' understanding of and proficiency with uncertainty through a range of perspectives with a range of intended goals~\cite[][references therein]{holmes_instructional_2023_v}. 

The most prominent physics education research thread has been through the classification of students' reasoning about uncertainty as either part of a point paradigm or set paradigm~\cite[e.g.,][]{buffler_development_2001}. Reasoning with the point paradigm includes ideas such as that any individual measurement could be exactly the ``true'' value, repeated measurements are not necessary and do not need to be combined, and measurements do not need to be listed with their uncertainties. Reasoning with the set paradigm includes ideas such as that any measurement is just an approximation of the phenomenon being measured, deviation between measurements is to be expected, combining repeated measurements helps establish the best estimate and its uncertainty, and all measurements should be reported with their uncertainties. 

Research characterizing students according to these two paradigms has generally found that students may exhibit either style of thinking depending on the question~\cite{lubben_point_2001, buffler_development_2001, allie_point_2001, lubben_experimental_2010}. For example, in these studies, many students studied exhibited point-like thinking on questions about whether repeated measurements were necessary. When comparing data sets, however, many students used mixed reasoning, which includes both point-like and set-like ideas. More recent work has found evidence that purely point-like reasoning is quite rare amongst introductory college physics students~\cite{pollard_impact_2020, kung_analyzing_2002, volkwyn_impact_2008, kung_university_2006}. Altogether, these findings suggest that student thinking about uncertainty is not unidimensional, may be context-dependent~\cite{leach_survey_1998},  and may not neatly fit into one of the two paradigms.

Evaluation of students as set- or point-like thinkers have evaluated students' \emph{procedural} knowledge about uncertainty, defined as being ``concerned with `doing science' \ldots rather than with the scientific concepts themselves. Thus, procedural knowledge (in the context of experimental work) will inform decisions, for example, when planning experimental investigations, processing data and using data to support conclusions''~\cite[][p. 1137]{buffler_development_2001}. For example, the Physics Measurement Questionnaire~\cite{allie_first-year_1998} presents students with experimental scenarios and sample measurement data and asks students about possible next procedures, including whether to perform repeated measurements, how to report a best estimate of the measurand, or how to compare pairs of measurements. Underlying these procedural decisions is a conceptual understanding of what uncertainty is and where it is coming from. In our companion work, we studied students' conceptions of where measurement uncertainty may be coming from in a range of experiments~\cite{Stump_TBD}. We find that students hold a wide range of ideas about the sources of uncertainty that depend on the experimental scenario. We also find that there is much more nuance in student thinking about measurement beyond the point and set paradigms.

\begin{figure*}[th]
  \subfloat[Projectile Motion scenario]{\fbox{\includegraphics[width=.6\textwidth]{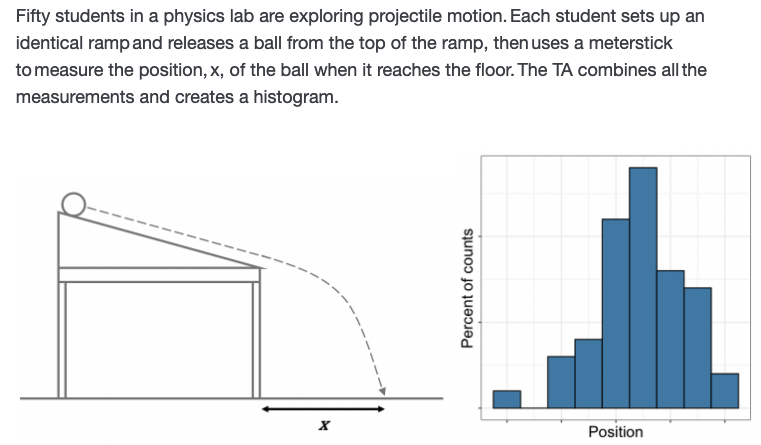}}\label{fig:PM}}\\
  \centering
  \subfloat[Stern-Gerlach scenario]{\fbox{\includegraphics[width=.37\textwidth]{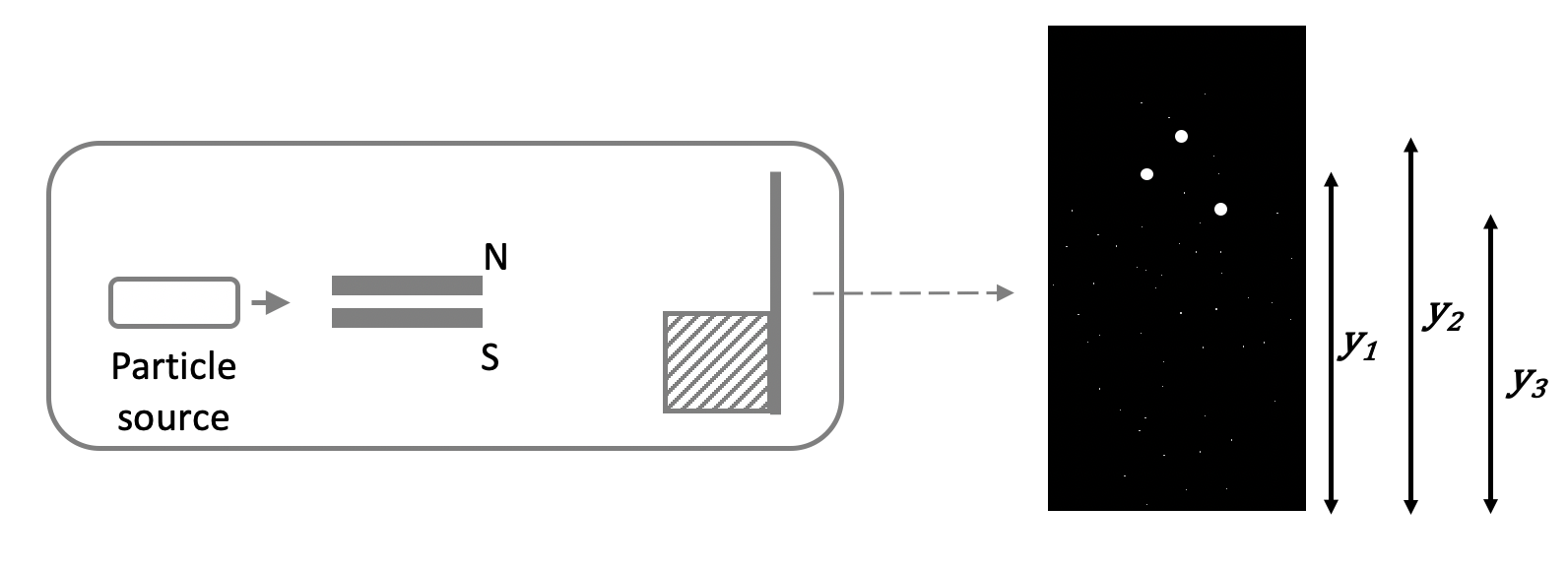}}\label{fig:SG}}\hspace{1em}
  \subfloat[Brownian Motion scenario]{\fbox{\includegraphics[width=.245\textwidth]{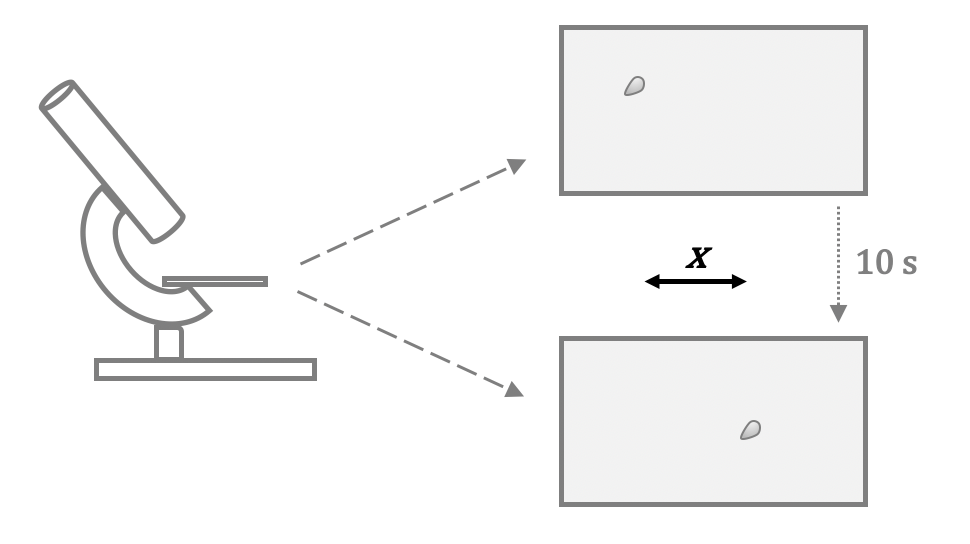}}\label{fig:BM}}\hspace{1em}
  \subfloat[Single-Slit scenario]{\fbox{\includegraphics[width=.285\textwidth]{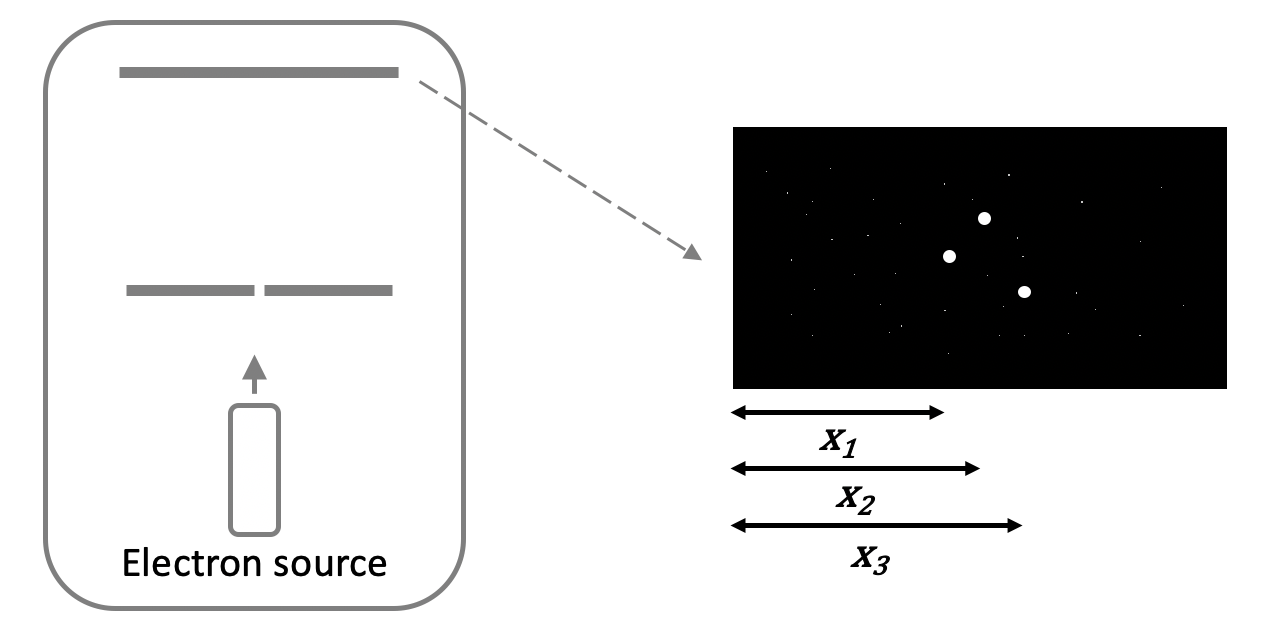}}\label{fig:SS}}
  \centering
    \caption{Diagrams for the four experimental scenarios included in the survey. The descriptive text and histogram for the Projectile Motion experiment are included as an example.}
    \label{fig:scenarios}
\end{figure*}
Altogether, these studies motivate a broader examination of student thinking about measurement uncertainty from new and distinct perspectives. This study is part of a broader project to evaluate these new perspectives. While previous assessments have asked students about what they think should be done next in an investigation, here we probe what students think will happen to experimental measurement data under two experimental interventions: 1. Adding additional data (under the same conditions) and 2. Obtaining new data by experts with the best possible equipment. We refer to these as the ``more'' and ``better'' data questions, respectively. We can contrast these questions to the procedural assessments of students' thinking about uncertainty, which ask students about \emph{whether} to take more data or improve measurement equipment. Those assessments infer point-like thinking through responses that indicate, for example, not needing additional data~\cite{lubben_point_2001}. Conversely, our questions can assess students' point- or set-like thinking through their perspectives of what \emph{will} happen under these experimental settings. For example, point-like thinking may be exhibited by students indicating that either setting will result in a single value. Set-like thinking, however, may be represented by any number of options (the distribution of data points may not change, may become more narrow but still have some variability, or may become more wide). Compared with the procedural assessments used to characterize point- and set-like thinking, students' justifications for their predictions will provide new and unique insights into their understanding of what is causing the variability and whether variability is a relevant and necessary construct for the experiments.

In addition to the new perspective provided by the types of questions, we also probe student thinking in several different experimental scenarios (Fig.~\ref{fig:scenarios}) that span both physics paradigm (classical and quantum mechanics) as well as theoretical expected outcome (single-value and a distribution). In contrast, previous assessments have only probed student thinking within a single-value classical mechanics scenario, which we include as one of our four scenarios. We also survey advanced physics students, as opposed to those in introductory laboratory courses. Our preliminary research has identified that students may consider classical and quantum systems as distinct experimental scenarios with different rules regarding uncertainty and variability~\cite{stein_student_2020, stump_student_2020, white_student_2020}. This distinction is not surprising given the documented examples of conflicts and tensions in student thinking between classical and quantum mechanics more generally~\cite{baily_development_2009, steinberg_influence_1999, marshman_framework_2015, krijtenburg-lewerissa_insights_2017}.

Ultimately, our research question is: How do students evaluate the impact of more and better data on experimental measurements across a range of experimental scenarios? In what follows, we find four main results. First, very few students exhibit definitional point-like thinking, where the measurements would result in a single value. Second, many students appropriately understand the ways in which more data do not impact the width of a distribution of measurements and the ways in which better data cause a distribution to narrow. Third, we find two key alternative concepts: a misattributed ``more data is better'' heuristic (or conflating standard deviation and standard error); and that fundamental physical principles (such as the Heisenberg Uncertainty principle) eclipse measurement limitations for some experimental scenarios. Finally, we find that student thinking about the impacts of more and better data vary marginally with experimental physics scenario.

\begin{figure*}[th]
\includegraphics[width=0.8\textwidth]{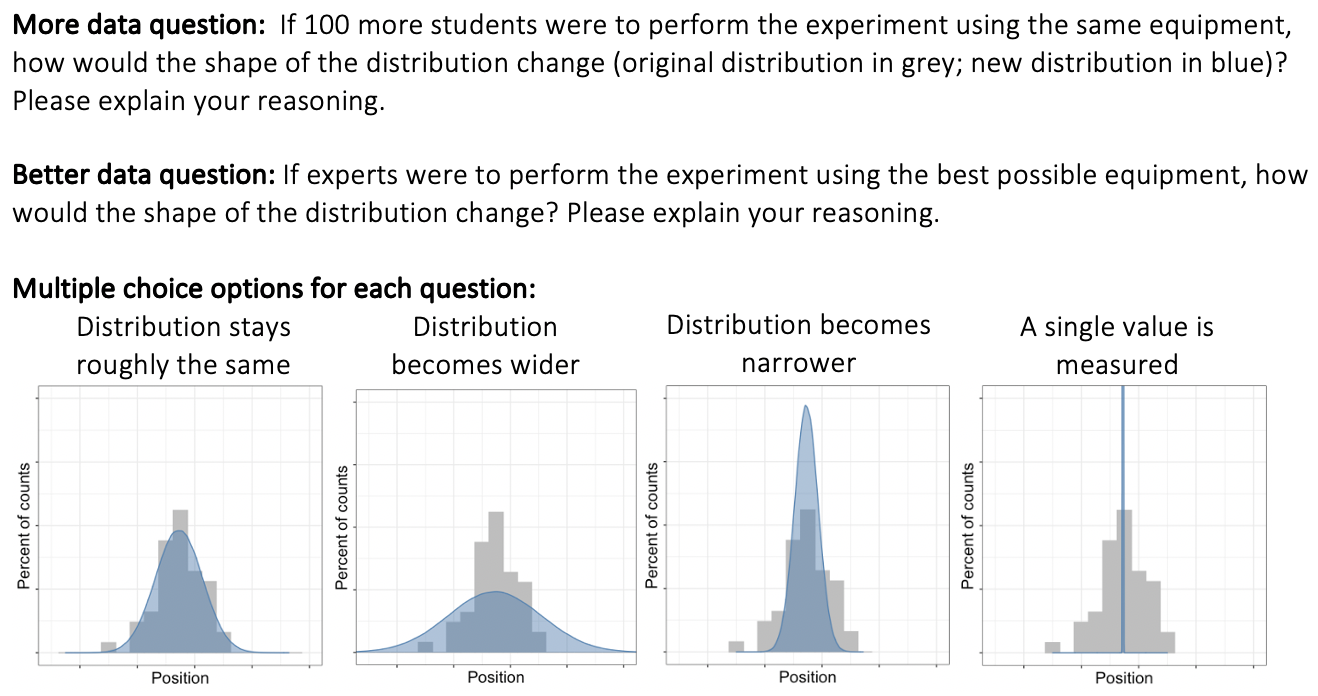}
\caption{More and better data questions with the multiple choice options.  After each question students were provided with a text box to explain their reasoning.}
\label{fig:questions}
\end{figure*}

\section{Methods}
This research is part of a larger project investigating student thinking about uncertainty and measurement. Data for this work come from a survey we developed for broad dissemination to probe student thinking across multiple perspectives. For analysis of student thinking about uncertainty from a different perspective, please see our companion paper~\cite{Stump_TBD}.

\subsection{Survey Development}

The survey centers around four experimental scenarios in which students in a laboratory course perform an experiment and take data.  Respondents are provided with a description of each scenario, a schematic of the experimental setup, and a histogram of fictitious data, as seen in Fig.~\ref{fig:scenarios}.  After the description of the experimental setup, respondents are asked a series of questions, after which they are presented with a second experimental scenario and asked the same set of questions again in this new scenario.  The first scenario shown to all students in this study was the Projectile Motion scenario (shown in Fig.~\ref{fig:PM}).  The second scenario was randomly chosen between a Stern-Gerlach experiment (Fig.~\ref{fig:SG}), a Brownian Motion experiment (Fig.~\ref{fig:BM}), and a single-particle, single-slit experiment (Fig.~\ref{fig:SS}).  These scenarios provide a set of conditions in a 2x2 of physics paradigm (classical: Projectile Motion and Brownian Motion; and quantum: Stern-Gerlach and Single-Slit) and theoretical expected outcome (single value: Projectile Motion and Stern-Gerlach; and distribution: Brownian Motion and Single-Slit). The Stern-Gerlach experiment has one of the two output channels blocked so that the theoretical experiment outcome would be a single value. The histogram of fictitious data is identical across scenarios and represents distance measured with a ruler.

The final two questions asked in each scenario are the focus of this paper.  These questions asked respondents what might happen to the distribution if {\it more} data or {\it better} data were taken. The full language of these questions can be seen in Fig.~\ref{fig:questions}. Respondents could select one option from the four presented in Fig.~\ref{fig:questions} and were given an open text box to explain their reasoning. These questions were designed to probe the ideas of set- and point-like thinking in more detail and through new perspectives. While the PMQ probes set and point-like thinking through procedural questions, such as ``what should you do next?'',  our questions are more conceptual in nature and instead provide students with the opportunity to describe how (if at all) they believe the results of an experiment would change if more data or better data were taken. For example, we would expect that point-like thinkers would select that the distribution would become a single value with either {\it more} or {\it better} data.

\subsection{Survey Distribution}

The survey was distributed to students in quantum mechanics courses at five institutions: Cornell University, University of St. Andrews, Michigan State University, University of Colorado Boulder, and California State University, Fullerton. The surveys were administered during the second half of both the Fall 2020 and Spring 2021 semesters electronically using Qualtrics.  Students were invited to participate by their course instructors and could opt to enter a draw to win a \$25 gift card if they responded to the survey.  In total, we received 150 completed student responses to the survey with the majority of students in the third year or fourth of their degree. The demographic information on the participating students is provided in Table~\ref{ta:demographics}. We do not have any information about the students' relevant prior knowledge, the ways their instruction had discussed measurement and uncertainty, nor whether they had completed experiments similar to the ones in this survey. We did ask students to indicate their level of comfort with each of the scenarios they saw in the survey. Over 90\% of student respondents reported being comfortable with the projectile motion and single-slit experiments and approximately 75\% of student respondents reported being comfortable with the Stern-Gerlach and Brownian motion experiments. Future work will seek to evaluate the ways prior knowledge and instruction interacts with students' reasoning on these and other questions about measurement and uncertainty.

\subsection{Data Analysis}
The questions analyzed in this work are multiple choice with a prompt to provide an explanation.  Student explanations were coded using a coding scheme developed during analysis of other questions on the survey and using the Modeling Framework for Experimental Physics~\cite{zwickl_model-based_2015}.  The coding scheme is described below and detailed information on its development and modifications for other uses can be found in Ref.~\cite{Stump_TBD}.

Each student response was given one or more of the following codes developed {\it a priori}: principles, limitations, statistics, or other. The first two codes were based on components of the Modeling Framework for Experimental Physics~\cite{zwickl_model-based_2015}. Examples of student explanations that fit each code and code definitions are provided in Table~\ref{tab:coding_scheme}. The {\it principles} code identifies reasoning that indicates the shape of the distribution is caused by variability inherent to a theoretical abstraction of the experiment.  Student reasoning about principles can be either classical or quantum mechanical in nature and can relate to the physical or measurement systems of the experiment, as distinguished in the Modeling Framework for Experimental Physics~\cite{zwickl_model-based_2015}.  The {\it limitations} code identifies reasoning that describes some practical limitations with the experiment, including instrumental imperfections and human error, or anything that is not inherent to the principles of the system being measured~\cite{zwickl_model-based_2015}.  The {\it statistics} code is unlike the other two in that it is not based on the modeling framework and is unique to the more and better data questions. This code is used when the student explains the shape of the new distribution through a statistical or data-driven lens, such as considering the statistical effects of adding more measurements to a data set.  These responses do not reference any element of the experimental scenario or physical or measurement systems, instead considering the statistical effects independently from what might be physically occurring. As in the examples in Table~\ref{tab:coding_scheme}, responses coded for statistics could be applied to any of the experiments. Students may have been considering how the probabilistic nature of the a particular system is the cause of the distributions they mention, but, as written, we do not have evidence to this effect. Finally, the {\it other} code was used when the explanation did not fit into any of the three previous categories, often because it was too vague or simply descriptive of their answer choice without explaining why they chose it. 

It is important to note that there is not necessarily a single correct code for any of the scenarios or questions.  Student responses assigned to each code demonstrate a wide range of expert and novice-like thinking.  Moreover, a student response that is coded as `limitations' (for example) does not indicate any lack of understanding or awareness of reasoning about the principles of the experiment.  Rather, the response only tells us what they chose to write to justify their multiple-choice answer to this particular question.  This is particularly true in the case of the statistics code, which is distinctly different from the other two in that the student is responding to the question by only discussing the data and not the experimental parameters (be it the physical or measurement system principles or limitations~\cite{zwickl_model-based_2015}. Both the answers students selected (single value, more narrow, same, and more wide) and the code assigned to their reasoning provide insights into student thinking. We find that the most information is learned when we look at the answer and reasoning codes together.

\begin{table*}[tb] 
  \caption{Coding scheme for the more and better data questions. Responses that did not fit into these codes or were too vague to code were coded as `other'.}
  \label{tab:coding_scheme}
  \begin{ruledtabular}
    \begin{tabular}{llp{0.6\textwidth}}
    \textbf{Code} & \textbf{Definition}& \textbf{Examples}\\
      \hline
   
    Principles & \multirow{4}{2 in}{Reasoning that indicates the shape of the distribution is caused by variability inherent to a theoretical abstraction of the experiment. It may be classical or quantum in nature.} & \textit{``As you increase the number of iterations your system should converge to the proper solution which should be one solveable value for a kinematic problem.''} (Projectile Motion) \\
    && \textit{``The distribution of values obtained comes from the fundamental randomness of the superposition breaking into one of the spin eigenstates. It doesn't have to do with equipment and expertise.''} (Stern-Gerlach) \\
    && \textit{ ``I think the randomness of Brownian motion is probably contributing more to the variation than experimental errors, so the distribution might stay the same regardless.''} (Brownian Motion)\\
    && \textit{ ``I think the dominant defining characteristic of the distribution remains quantum mechanical and inherent to the physics of the experiment, and so the general of the shape of the distribution would stay the same, but become more perfect to a particular functional form.''} (Single-Slit)\\
     \hline 

    Limitations & \multirow{5}{2 in}{Reasoning that describes some practical limitations with the experiment, including instrumental imperfections to human error, that is not inherent to the principles of the physical or measurement system.}  & \textit{``Using the best possible equipment means reduction in human error (like giving initial push and initial position on ramp).''} (Projectile Motion)\\
    && \textit{``The expert will eliminate the difference in the initial velocity, so the position uncertainty will be reduced to zero, that is, all the upwardly deflected particles will reach the same position.''} (Stern-Gerlach)\\
    &&\textit{``I know some labs are set on material that absorbs vibrations and you can probably compensate somewhat for the lingering motion from moving the plate.''} (Brownian Motion)\\
    &&\textit{``The better equipment will allow more accuracy in electron production and precision in measurements. Also any irregularities in the slit can be reduced.''} (Single-Slit)\\
    \hline

    Statistics & \multirow{4}{2 in}{Reasoning that explains the shape of the new distribution through a statistical or data-driven lens, such as considering the statistical effects of adding more measurements to a data set.} & \textit{``By increasing the sample size, were likely to get more measurements around the peak of the original histogram and less outlying measurements. This is because an increase in sample size decreases the standard deviation for measurement.''} (Projectile Motion)\\
    && \textit{``The expectation value (or peak) should remain the same under multiple trials however the probability to measure each thing should also stay the same.''} (Stern-Gerlach)\\
    &&\textit{``The underlying distribution is gaussian so you expect it to remain about the same.''} (Brownian Motion)\\
    &&\textit{``The positions will be different for the 3 electrons for any 3 particular measurements but the more measurements that are taken, the more the statistical shape of the position probability density will begin to take shape. ''} (Single-Slit)\\
   
    \end{tabular}
  \end{ruledtabular}
\end{table*}

A subset of the data (95 responses) was coded by three of the authors. Cohen's kappa was calculated to be greater than 0.8 across each pair, indicating fair inter-rater reliability.  All disagreements were then discussed and resolved, with clarifications then made to the codebook to reduce the occurrence of further disagreements.  The full dataset was then coded by the first author and a fourth author read through all responses and associated codes to double check the consistency with which codes were applied to responses. Fewer than a handful of codes were changes during this final review.

Our goal here was to uncover overarching trends in the student responses. We chose not to perform statistical tests because several of the proportions were too small to perform $\chi^2$ tests of distinguishability and relying on $p$-values for interpretation is strongly misleading~\cite[see, for example, ][]{cohen_earth_1994, cumming_new_2013, nosek_preregistration_2018}. Any statements about ``differences'' or ``similarities'' in our results section, therefore, should be interpreted as ``qualitatively distinguishable from the graph'' or ``not qualitatively distinguishable from the graph'', respectively.

\section{Results}
We break down our discussion of the results by first looking at which multiple-choice option students selected to both the more and better data questions for each experiment (Section~\ref{sec:res_mc}). We then look more closely at the reasoning students provided to justify their answers (Section~\ref{sec:res_reason}).

\subsection{Student responses to how the distribution might change}\label{sec:res_mc}

\begin{figure*}[t]
    \centering
    \includegraphics[width=0.8\textwidth]{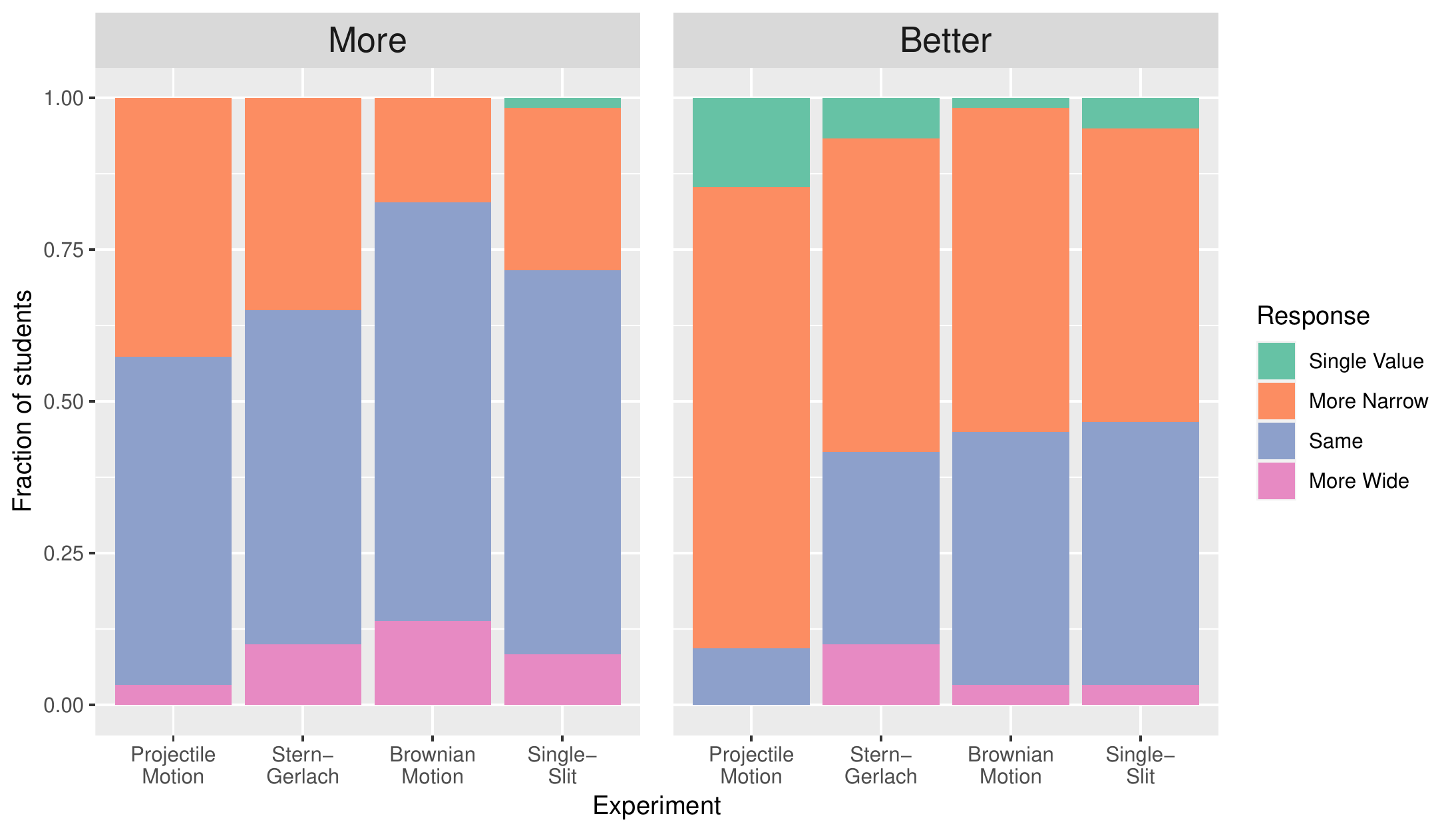}
    \caption{Stacked bar plot of student responses to the more and better data questions for each experimental scenario.  }
    \label{fig:stackedbar}
\end{figure*}

The distribution of student responses for each question and each experimental scenario are shown in Figure~\ref{fig:stackedbar}. In response to both questions and all scenarios, we see that most students selected that the distribution would get more narrow or stay the same, with a small number of students selecting that it would become more wide or result in a single value.  

Looking more closely at the `more data' responses, we note almost no students indicated that more data would result in a single value. This is perhaps unsurprising, as the only change to the experiment is taking additional data and the initial measurements resulted in a distribution. We also notice a few students indicating that the distribution will become more wide, with more students selecting this option for the three experiments that are either quantum mechanical and/or have a theoretically expected distribution (that is, not the Projectile Motion experiment). 

The most common response for how the distribution might change when more data are taken is that the distribution will stay the same. We found this to be the case for all four experimental scenarios. We identify this response as the most expert-like response, as more data collected with the same procedures and equipment will not affect the overall variability in the measurements, though the estimate of the mean of those data would have smaller uncertainty. The next most common response is that the distribution would become more narrow. The fewest number of students selected this option in the Brownian Motion scenario.

In response to the `better data' question, we notice a larger difference between experiments. We see that the Projectile Motion scenario receives a much larger proportion of more narrow responses than the other experiments. We identify this as an 'expert-like' response (across all experiments) because experts with the best equipment will perform measurements with fewer experimental limitations than students with basic equipment. 

We do not notice much difference between the other three experiments with students expecting the distributions to become more narrow (most frequent response) or stay the same (second most frequent response) at similar rates. Additionally, the Projectile Motion experiment received the largest number of responses indicating that there would be a single value, and students were overall more likely to predict that a single value would result from better data than from more data.

\subsection{Student explanations about what happens to the distributions} \label{sec:res_reason}

We now take a look at the explanations students provided to justify their choices about the distributions. Because most students indicated that the distributions would either become more narrow or stay the same, we only investigate explanation codes for students who selected these two options. The distribution of student responses for these two answers are provided in the scaled pie charts in Figs.~\ref{fig: bb100} and \ref{fig: bbe}.

\subsubsection{`More data' explanations}
\begin{figure*}[t]
    \centering
    \includegraphics[width=\textwidth]{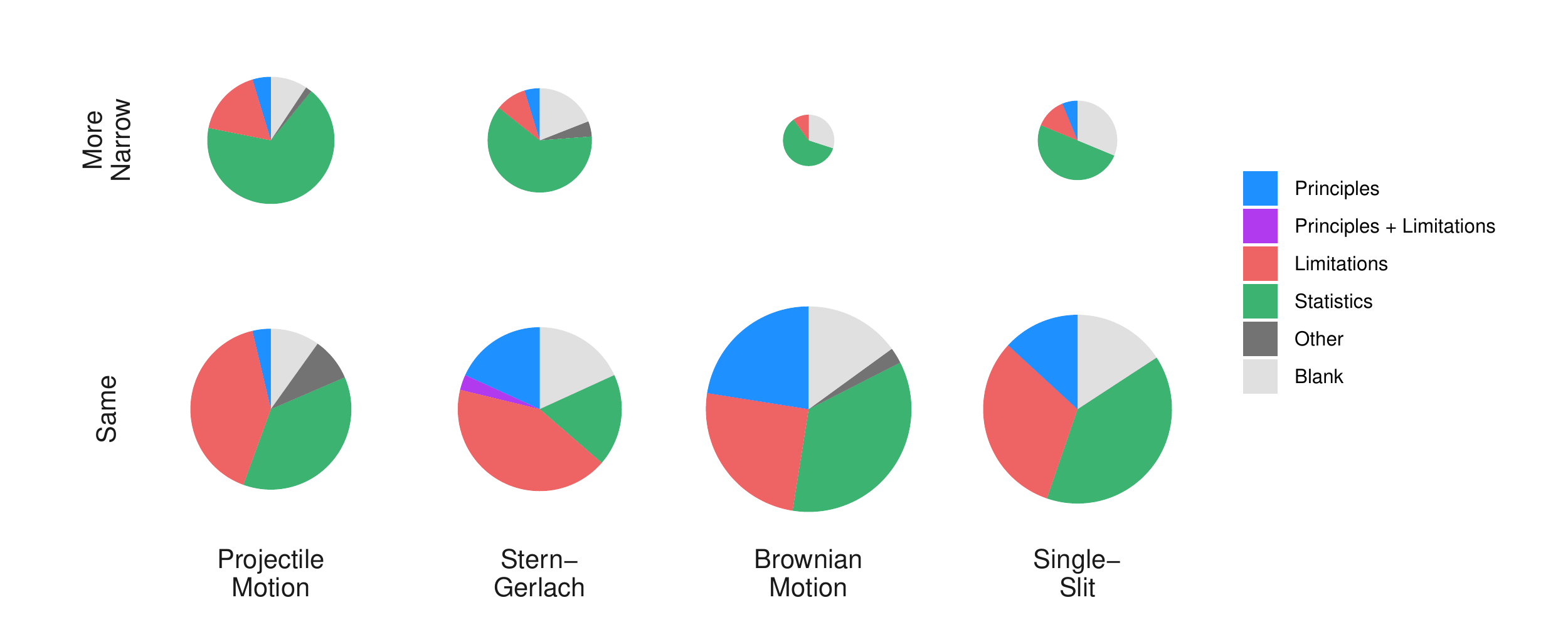}
    \caption{Scaled pie charts of the more data explanation codes for more narrow and same distribution responses. The pie charts are normalized vertically according to the proportion of students who selected that distribution and the slices show the proportional breakdown of explanation codes.}
    \label{fig: bb100}
\end{figure*}  
From Fig. \ref{fig: bb100}, we see that students who indicated the distribution would become more narrow used different reasoning than those who indicated the distribution would stay the same. When students answered more narrow, their explanations were primarily coded for Statistics. These responses referred to purely statistical effects that do not consider the physical system, such as ``the law of large numbers'' or saying that more data reduces the standard deviation. For example, one student considering the Projectile Motion scenario said: ``Distribution will indeed be a Gaussian, as you increase the number of trials, the random uncertainty in the outcome will decrease, standard deviation will decrease and hence standard error decrease to create narrower distribution." Another student considering the Brownian Motion scenario said: ``If there is a true value of x, then distribution would become narrower because of how most data points should trend to the true value of x. There will still be less points on the sides." A minority of students used limitations to explain the distribution becoming more narrow, typically inferring that more data reduces student-driven variability. For example, a student considering the Single-Slit scenario said ``If more students conduct the experiment, then the distribution will become more narrow because of the uncertainty caused by human-error is reduced.'' Another student considering the Projectile Motion scenario said: ``I want to say the distribution becomes narrower, because the more that are set up the same, there are less factors to change the distribution and therefore the percentage of counts increases at the average.''

Most students, however, indicated the distribution would stay the same, with a range of explanations across scenarios. Many students also used purely statistical ideas to explain the distribution staying the same, typically describing that more data would not change the standard deviation. For example, one student considering the Stern-Gerlach scenario said: ``The distribution may become smoother, but will not change in characteristic shape as more data is collected with the same.'' Another student considering the Brownian Motion experiment said: ``Increasing n should not change the distribution since n is already large.''

Similar numbers of students, however, used limitations to explain the distribution staying the same, such as that the additional students were still using the same equipment and procedures and so the same sources of uncertainty exist. For example, one student considering the Projectile Motion scenario said: ``A hundred students will make as many random errors as fifty - the standard deviation of the measurement shouldn't change.'' Another student considering the Brownian Motion scenario said: ``Since the equipment is the same, the systematic errors should be similar. Presumably the other variables are the same (Temp), so statistical errors are similar as well. Thus the distribution is similar.''

A large minority of students also argued for physical and measurement principles, particularly for the Stern-Gerlach, Brownian Motion, and Single-Slit experiments. In these cases, students primarily indicated that the phenomena themselves were random and so additional data would not remove that inherent randomness. For example, a student considering the Stern-Gerlach scenario said: ``The limitation is not the students' accuracy or the sample size but the quantum mechanical system.'' Another student considering the Brownian Motion experiment said: ``The motion is random, so it does not matter how many trials are taken.''

Interestingly, we see relatively little variation between experimental scenarios on the explanation codes. Students used similar reasoning to explain why the distribution would become more narrow regardless of the physics paradigm or theoretical expected outcome. For the distribution staying the same, the relative proportions differ in only minor ways (such as fewer principles explanations for Projectile Motion, and more limitations codes and fewer statistics codes for Stern-Gerlach). In addition, the role of statistics and limitations appear to be context independent, in that many of the quoted responses above could have been associated with any of the experimental scenarios.

\subsubsection{`Better data' explanations}

\begin{figure*}[t]
    \centering
    \includegraphics[width=\textwidth]{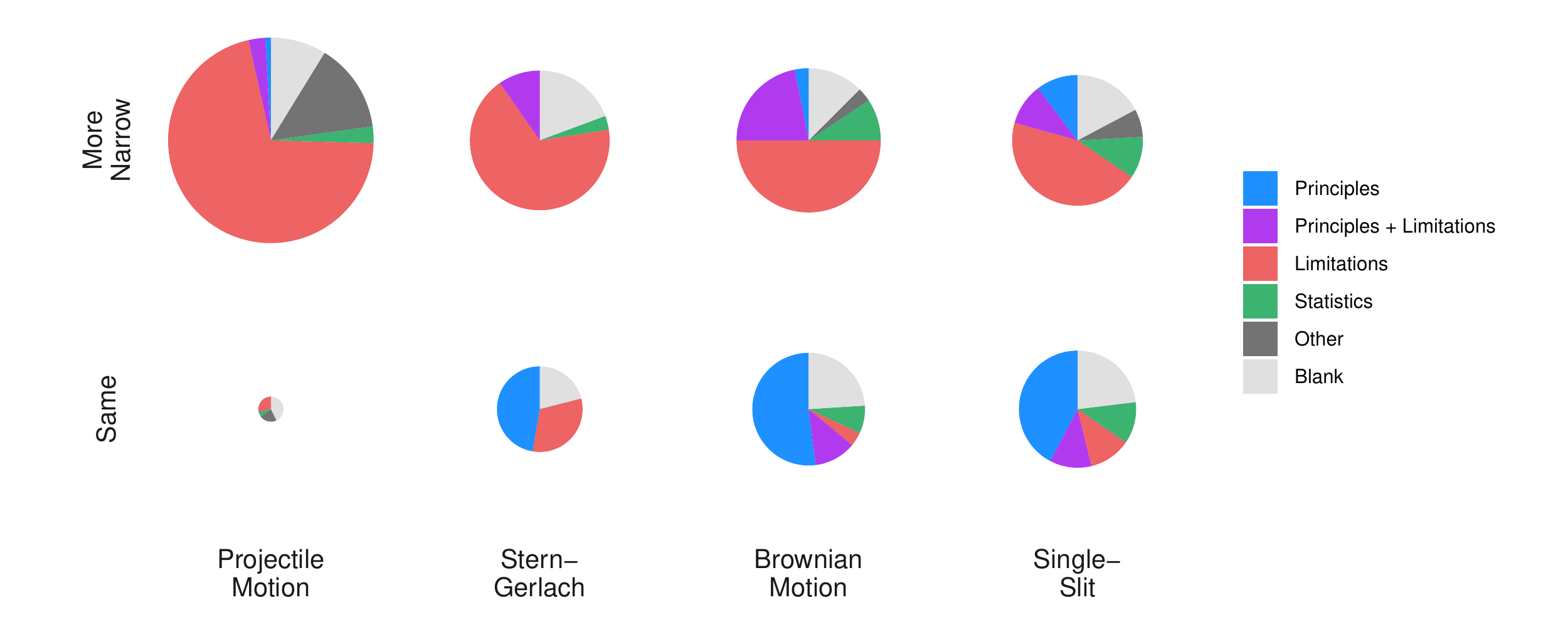}
    \caption{Scaled pie charts of the better data explanation codes for more narrow and same distribution responses. The pie charts are normalized vertically according to the proportion of students who selected that distribution and the slices show the proportional breakdown of explanation codes.}
    \label{fig: bbe}
\end{figure*}
From Fig. \ref{fig: bbe}, we see that when students indicated the distribution would become more narrow, their explanations were primarily coded for limitations. Most often, students' responses related to the fact that experts with expert equipment would remove much of the variability, uncertainty, and error present in the student measurements. For example, a student considering the Projectile Motion scenario said: ``The distribution would become narrower because, in my opinion, it is safe to assume that the experts in this case will use significantly better equipment that has much lower systematic error associated with it. This corresponds to a narrower distribution!'' Another student considering the Single-Slit scenario said: ``Considering that I think the variability results more so from the quality of the equipment, perhaps the best possible equipment that the experts are using may result in a narrower and more precise distribution.'' Many students also used limitations to justify why the distribution would not be a single value. For example, a student considering the Projectile Motion scenario said: ``The best possible equipment will still have error associated with it, not possible to reduce the error to zero, will always have random fluctuations. Hence, the distribution will still be a Gaussian and with more measurements taken, would become narrower.''

Similarly, an interesting minority of students used principles when explaining the distribution becoming more narrow (most commonly alongside limitations). These principles responses particularly referred to justifying the choice of the distribution becoming more narrow rather than a single value. For example, one student considering the Projectile Motion scenario said: ``This minimized a few sources of uncertainty, but there are still some that can't be erased completely. Physics is never completely deterministic or perfect." Another student considering the Brownian Motion scenario said: ``Experts can reduce some of the sources of the distribution but not others so that would make the distribution narrower. However, I think the primary source of the distribution (random Brownian motion) cannot be reduced so the distribution would not be narrowed as much as above.'' We also consider such responses to be 'expert-like' because the students appropriately identify that these systems have an inherently limiting uncertainty. 

For the smaller fraction who indicated the distribution would stay the same, many argued using principles. As with the more data explanations, students argued that, for the Stern-Gerlach, Brownian Motion, and Single-Slit distributions, the variability in the system was inherent to the physical system and could not be removed, even by experts with expert equipment. For example, a student considering the Stern-Gerlach scenario said: ``The imprecise measurement doesn't play a role here. The distribution of values obtained comes from the fundamental randomness of the superposition breaking into one of the spin eigenstates. It doesn't have to do with equipment and expertise.'' Another student considering the Single-Slit scenario said: ``The thing here is, that because of quantum randomness -- it should stay the same!''

We again see relatively little variation between experiments on the explanation codes, though projectile motion appears to be distinct from the other three scenarios. For students who indicated the distribution would become more narrow, a lower proportion of students used principles explanations for projectile motion as compared to the Stern-Gerlach, Brownian Motion, and Single-Slit experiments (which are otherwise similar in proportions). For students who indicated the distribution would stay the same, we see minor differences in the relative proportions of statistics and limitations codes among the Stern-Gerlach, Brownian Motion, and Single-Slit experiments, with the major reasoning being tied to inherent principles of the physical systems. Because so few students indicated that the distribution for the Projectile Motion experiment would stay the same, we can say little about their explanations.

\section{Discussion}

In this study, we evaluated students' perceptions of what would happen to hypothetical distributions of measurements from four different physics experiments under two settings: collecting `more data' (100 more students with the same equipment) or collecting `better data' (experts with the best possible equipment). We identified whether students thought the distributions would become wider, stay the same, become more narrow, or result in a single value and evaluated their reasoning for their choice. 

A primary motivation for this analysis was to shed new light on previous understandings of student thinking about measurement; namely, the point and set paradigms~\cite[e.g.,][]{buffler_development_2001}, which classify students in terms of either thinking about experimental results as individual measurements or a distribution of measurements. Our expectation was that students exhibiting set-like thinking would select that any of the distributions would either become more narrow, stay the same, or become wider with `more' or `better' data. We similarly expected that students exhibiting point-like thinking would select that the distributions would result in a single value (represented by a delta function) in response to the `better data' question. In our data, very few students indicated the distribution would result in a single value: around 10\% in the `better data' question for the Projectile Motion experiment, fewer than 5\% in the `better data' question for the other experiments, and effectively 0\% in the `more data' questions across experiments.

This result inspires multiple plausible explanations. First, perhaps very few students in our sample were exclusively point-like thinkers. This explanation is supported by work across several institutions that claim that exclusive point-like thinking, as measured by the Physics Measurement Questionnaire, is quite rare~\cite{volkwyn_impact_2008,kung_university_2006, pollard_impact_2020, lippmann_students_2003}. Second, point-like thinking may be more nuanced than expecting that one can measure exactly the ``true value'' under ideal conditions. Many students who thought the distributions would become more narrow explicitly commented on the distribution centering on the true value, but qualified their answer with the idea that uncertainty and errors could never be completely eliminated. Given that point-like thinking focuses on true values and set-like thinking focuses on distributions~\cite{buffler_development_2001}, one might argue that this line of reasoning could be consistent with either type of thinking. We argue here, therefore, that our survey questions (and their associated responses) provide a new perspective on what it means to be a point- or a set-like thinker. 

More in line with set-like thinking, the majority of students across questions and experiments indicated the distributions would either stay the same or become more narrow, but the relative proportions differed for the `more' and `better' data questions. The explanations for their choices also differed between the `more' and `better' data questions, as well as between experimental scenarios.

On the `more' data question, most students appropriately expected the distribution would stay the same, with the primary justification being that the additional data did not affect the limitations of the experiment or that additional data would not change the standard deviation. A minority of students justified this response with the inherent principles of the physical or measurement processes, although this explanation was much less frequent in the Projectile Motion scenario than the other experiments. 

In contrast, students who indicated the distribution would become more narrow primarily argued through purely statistical reasoning. We suggest that students were applying a ``more data is better'' heuristic or a form of a phenomenological primitive~\cite{disessa_toward_1993}. Previous work in statistics education has found that students often use analysis procedures as rote algorithms~\cite{bakker_lessons_2011}. Plausibly, the rote algorithm associated with collecting many data points may have been internalized through the heuristic ``more data is better.'' Under this heuristic, students appropriately drew on one of the key ideas behind set-like thinking: one needs multiple data points to estimate any phenomenon~\cite{buffler_development_2001} and, indeed, we become much more confident in our estimate of a parameter (e.g., the mean of the distribution) with many more data points. This idea may have become over-generalized, however, such that students inferred that the distribution itself must become more narrow. This explanation is particularly compelling given that the explanations coded for statistics used purely statistical reasoning; they were not considering the physical situation. In many cases, students also stated explicitly that the standard deviation would become smaller. Future work should evaluate whether students are making this claim through the heuristic described above or through short-circuited mathematical reasoning. For example, the division by $N$ in the equation for standard deviation may lead students to infer that a larger $N$ makes the standard deviation smaller (ignoring the added terms in the summation in the numerator). Alternatively, students may simply be confusing the standard deviation with the standard uncertainty of the mean (with its extra division by the square root of $N$).

On the `better' data question, most students appropriately expected the distribution would become more narrow, with almost all of the codeable responses reflecting an improvement in the limitations of the experiment. A small minority of students commented on the principles of the physical systems to justify the distribution becoming more narrow but not reaching a single value. For the smaller fraction of students who suggested the distributions would stay the same, the primary reasoning was the inherent principles of the physical or measurement processes in the Stern-Gerlach, Brownian Motion, and Single-Slit scenarios.

In contrast to our previous work~\cite{white_student_2020}, these data do not show a clear split between student thinking across classical (Projectile  Motion and Brownian Motion) versus quantum (Stern-Gerlach and Single-Slit) mechanical scenarios. Nor do the data show a clear split between single-value deterministic (Projectile Motion and Stern-Gerlach) and probabilistic distribution (Brownian Motion and Single-Slit) experiments. Instead, we see that students responded differently to the Projectile Motion experiment than to the other three experiments, such that reasoning amongst the other three experiments was quite similar. 

This result is non-trivial, particularly when considering the Stern-Gerlach scenario. Theoretically, one could reasonably evaluate the two single-value deterministic experiments (Projectile Motion and Stern-Gerlach) in similar ways. That is, based on physical principles alone, experts with expert equipment would approach a single value distribution for the Stern-Gerlach experiment in much the same was as in the Projectile Motion experiment. The reduction in limitations should make the distribution more narrow and, potentially, single valued. The existence of quantum mechanical principles do not justify the distribution staying the same for the Stern-Gerlach experiment and yet, a higher proportion of students expected the distribution would stay the same for the Stern-Gerlach experiment than for the Projectile Motion experiment, with either principles or limitations justifying that choice.

We infer, therefore, that many students are seeing a fundamental distinction between the principles of classical and quantum mechanics for single-valued experiments, but not for distribution experiments. Compared to the two single-value deterministic experiments, students considered the effect of `more' and `better' data on the two distribution experiments (Brownian Motion and Single-Slit) in similar ways. While previous research has suggested that students carry over much of their thinking about classical mechanics to quantum mechanics~\cite{baily_development_2009, steinberg_influence_1999, marshman_framework_2015, krijtenburg-lewerissa_insights_2017}, our results show that this carry-over may be context dependent. 

\subsection{Implications for instruction}
We argue that these results motivate two important lessons for instruction. First, from the `more' data results, the prevalence of the Statistics code as justification for the distribution becoming more narrow suggests that lab instruction should more explicitly address the applicability of the ``more data is better'' heuristic in multiple experimental scenarios. One might address the heuristic through a more careful consideration of the difference between standard deviation and standard uncertainty of the mean, with explicit learning goals and activities focused on this distinction. Such instruction should attend to the productiveness of the ``more data is better'' heuristic (building on students' existing and productive resources~\cite[e.g.,][]{hammer_student_2000, robertson_resources-oriented_2022}), while distinguishing the uncertainty in individual measurements from the uncertainty in the mean of multiple measurements. Given that the heuristic was equally present across experimental scenarios, a purely statistical treatment may be sufficient. Our companion work evaluating students' characterizations of the sources of uncertainty~\cite{Stump_TBD}, however, motivates the need for instruction that links this heuristic to the physical and measurement properties of the experiment as well, such as by encouraging students to consider the sources of uncertainty and whether more data will reduce the effect of those sources. This treatment is supported by the large proportion of students already attending to limitations of the experiment in arguing for how more data does not change the distribution. 

Second, from the `better' data results, the prevalence of the Principles code as justification for the distribution staying the same suggests that theoretical quantum mechanics instruction should more explicitly address the experimental limitations in quantum mechanical measurement as distinct from the uncertainty resulting from the principles of quantum mechanics. This recommendation is further supported by our companion work evaluating students' characterizations of the sources of uncertainty~\cite{Stump_TBD}, which similarly finds that students strongly attend to the quantum mechanical effects of the Stern-Gerlach experiment, even when they do not explain the observed variability from the experiment. For the students who expected the distributions to stay they same, they largely did so based on physical principles with little to no consideration for the limitations of the experiments. While an experiment may reach a ``quantum limit'', students should be provided with experience in analyzing when that limit might be reached.

\subsection{Implications for future work}

This study includes multiple limitations and open questions that inspire further study. As with any physics education research study, our finite sample size motivates replication studies with larger data sets, particularly recruiting from primarily undergraduate institutions and minority-serving institutions. Data should also be collected from introductory and advanced students for the Projectile Motion scenario to understand the ways in which student thinking about that scenario may evolve over time. Given the high reliability of the coding scheme, we are intrigued by the possibility of natural language processing to parse the student explanations to facilitate much larger data sets, which has been recently carried out using the Physics Measurement Questionnaire~\cite{wilson_classification_2022}. Future work should also consider additional experimental scenarios to evaluate the generalizability of our claims regarding the role of physics paradigm and theoretical expected outcome. For example, how might students evaluate the role of more or better data in various biological systems? 

Future work should also seek to disentangle why student reasoning about the Projectile Motion scenario is distinct from reasoning about the other scenarios. To that effect, one direction would be to simply collect more data (pun intended) from the upper-level scenarios (Brownian Motion, Single Slit, and Stern-Gerlach). Another direction would be to provide upper-division students with two of the upper-level scenarios and compare their reasoning directly. This analysis would test whether the differences observed for Projectile Motion are due to our having students complete the Projectile Motion scenario and one of the other scenarios, such that their responses to the second scenario are all comparative to the Projectile Motion scenario (and education research makes clear the power of contrasting cases~\cite[e.g.,][]{bransford_new_1989, schwartz_time_1998, schwartz_inventing_2004, chase_explaining_2010,schwartz_practicing_2011, chi_seeing_2012, roelle_effects_2015, heim_what_2022}. What would the results look like if students had not seen the Projectile Motion scenario? Future work could also test the robustness of this reasoning to other introductory-level classical experiments, such as pendulum motion or masses on springs.

Given our interpreted contrasts to student thinking about sources of uncertainty and the point and set paradigms, our future work will look at the relationships between student thinking on different types of uncertainty questions. For example, does a student who draws on limitations when considering the role of `more data' primarily list limitations as the sources of variability? Or does one survey question prompt different reasoning than another? Similarly, how do students' explanations regarding the role of `more' or `better' data compare to their reasoning on the Physics Measurement Questionnaire? With a larger dataset, we would be better able to breakdown student reasoning across these various categories. 

Finally, research should evaluate how instruction in both lab and theory courses impacts students' reasoning across these items and contexts. 

\acknowledgments{This material is based upon work supported by the National Science Foundation Grants No.~DUE-1808945 and No.~DUE-1809178 and the National Science Foundation Graduate Research Fellowship under Grant No. DGE-2139899. We are grateful to Courtney White for her initial work on this project, our project evaluator, Ben Zwickl, for fruitful discussions of this work and to Peter Lepage and the Cornell Physics Education Research Lab for comments and feedback on this work over the last four years.}

\bibliography{references.bib, ref2.bib}

\appendix*

\section{Raw data tables}

Participant demographic data is presented in full in Table~\ref{ta:demographics}. We also provide the data corresponding to the results figures in the main text in Tables~\ref{ta:raw1}, \ref{ta:raw2}, \ref{ta:raw3}. In each case, we include the total number of responses in each category, rather than percentages. 

\begin{table}[b] 
  \caption{Demographic information for the students considered in this analysis.}
  \label{ta:demographics}
  \begin{ruledtabular}
    \begin{tabular}{p{.35\textwidth}c}
    \textbf{Institution} & \textbf{\# of students}\\
    \quad California State University Fullerton & 8\\
    \quad Cornell University & 83\\
    \quad Michigan State University & 30\\
    \quad University of Colorado Boulder & 26\\
    \quad University of St. Andrews & 3\\
    \textbf{Year of college} \\
    \quad Second year (sophomore) & 9\\
    \quad Third year (junior) & 75\\
    \quad Fourth year + (senior) & 49\\
    \quad Graduate student & 11\\
    \quad Unspecified & 6\\
    \textbf{Gender}\\
    \quad Female & 40\\
    \quad Male & 104\\
    \quad Non-binary & 2\\
    \quad Unspecified & 4\\
    \textbf{Race/ethnicity}\\
    \quad American Indian or Alaska Native & 1\\
    \quad Asian or Asian American & 44 \\
    \quad Black or African American & 2\\
    \quad Hispanic or Latinx & 18 \\
    \quad Native Hawaiian or other Pacific Islander & 1\\
    \quad Prefer to self-describe & 4\\
    \quad White & 78\\
    \quad Unspecified & 16\\
    \end{tabular}
  \end{ruledtabular}
\end{table}

\begin{table}[]
\caption{Raw frequency data corresponding to Fig.~\ref{fig:stackedbar}.}\label{ta:raw1}
\begin{tabular}{lllr}
\hline\hline
Question & Scenario & Response & Frequency\\
\hline
More & Projectile Motion & Single Value & 0\\

More & Projectile Motion & More Narrow & 64\\

More & Projectile Motion & Same & 81\\

More & Projectile Motion & More Wide & 5\\
\hline
More & Stern- Gerlach & Single Value & 0\\

More & Stern- Gerlach & More Narrow & 21\\

More & Stern- Gerlach & Same & 33\\

More & Stern- Gerlach & More Wide & 6\\
\hline
More & Brownian Motion & Single Value & 0\\

More & Brownian Motion & More Narrow & 10\\

More & Brownian Motion & Same & 40\\

More & Brownian Motion & More Wide & 8\\
\hline
More & Single- Slit & Single Value & 1\\

More & Single- Slit & More Narrow & 16\\

More & Single- Slit & Same & 38\\

More & Single- Slit & More Wide & 5\\
\hline
Better & Projectile Motion & Single Value & 22\\

Better & Projectile Motion & More Narrow & 114\\

Better & Projectile Motion & Same & 14\\

Better & Projectile Motion & More Wide & 0\\
\hline
Better & Stern- Gerlach & Single Value & 4\\

Better & Stern- Gerlach & More Narrow & 31\\

Better & Stern- Gerlach & Same & 19\\

Better & Stern- Gerlach & More Wide & 6\\
\hline
Better & Brownian Motion & Single Value & 1\\

Better & Brownian Motion & More Narrow & 32\\

Better & Brownian Motion & Same & 25\\

Better & Brownian Motion & More Wide & 2\\
\hline
Better & Single- Slit & Single Value & 3\\

Better & Single- Slit & More Narrow & 29\\

Better & Single- Slit & Same & 26\\

Better & Single- Slit & More Wide & 2\\
\hline\hline
\end{tabular}
\end{table}

\begin{table*}
\caption{Raw frequency data associated with Fig.~\ref{fig: bb100}.}\label{ta:raw2}
\begin{tabular}{lllr}
\hline\hline
Scenario & Response & Code & Freq\\
\hline
Projectile Motion & More Narrow & Blank & 6\\

Projectile Motion & More Narrow & Statistics & 43\\

Projectile Motion & More Narrow & Principles & 3\\

Projectile Motion & More Narrow & Principles + Limitations & 0\\

Projectile Motion & More Narrow & Limitations & 11\\

Projectile Motion & More Narrow & Other & 1\\

Projectile Motion & More Wide & Blank & 1\\

Projectile Motion & More Wide & Statistics & 2\\

Projectile Motion & More Wide & Principles & 0\\

Projectile Motion & More Wide & Principles + Limitations & 0\\

Projectile Motion & More Wide & Limitations & 1\\

Projectile Motion & More Wide & Other & 1\\

Projectile Motion & Same & Principles & 3\\

Projectile Motion & Same & Blank & 8\\

Projectile Motion & Same & Principles + Limitations & 0\\

Projectile Motion & Same & Statistics & 30\\

Projectile Motion & Same & Other & 7\\

Projectile Motion & Same & Limitations & 33\\

Projectile Motion & Single Value & Principles & 0\\

Projectile Motion & Single Value & Blank & 0\\

Projectile Motion & Single Value & Principles + Limitations & 0\\

Projectile Motion & Single Value & Statistics & 0\\

Projectile Motion & Single Value & Other & 0\\

Projectile Motion & Single Value & Limitations & 0\\
\hline
Stern- Gerlach & More Narrow & Other & 1\\

Stern- Gerlach & More Narrow & Principles & 1\\

Stern- Gerlach & More Narrow & Blank & 4\\

Stern- Gerlach & More Narrow & Principles + Limitations & 0\\

Stern- Gerlach & More Narrow & Statistics & 13\\

Stern- Gerlach & More Narrow & Limitations & 2\\

Stern- Gerlach & More Wide & Other & 2\\

Stern- Gerlach & More Wide & Principles & 1\\

Stern- Gerlach & More Wide & Blank & 1\\

Stern- Gerlach & More Wide & Principles + Limitations & 0\\

Stern- Gerlach & More Wide & Statistics & 2\\

Stern- Gerlach & More Wide & Limitations & 0\\

Stern- Gerlach & Same & Other & 0\\

Stern- Gerlach & Same & Blank & 6\\

Stern- Gerlach & Same & Statistics & 6\\

Stern- Gerlach & Same & Principles & 6\\

Stern- Gerlach & Same & Principles + Limitations & 1\\

Stern- Gerlach & Same & Limitations & 14\\

Stern- Gerlach & Single Value & Other & 0\\

Stern- Gerlach & Single Value & Blank & 0\\

Stern- Gerlach & Single Value & Statistics & 0\\

Stern- Gerlach & Single Value & Principles & 0\\

Stern- Gerlach & Single Value & Principles + Limitations & 0\\

Stern- Gerlach & Single Value & Limitations & 0\\
\hline\hline
\end{tabular}
\quad
\begin{tabular}{lllr}
\hline\hline
Scenario & Response & Code & Freq\\
\hline
Brownian Motion & More Narrow & Statistics & 6\\

Brownian Motion & More Narrow & Other & 0\\

Brownian Motion & More Narrow & Blank & 3\\

Brownian Motion & More Narrow & Limitations & 1\\

Brownian Motion & More Narrow & Principles & 0\\

Brownian Motion & More Narrow & Principles + Limitations & 0\\

Brownian Motion & More Wide & Statistics & 0\\

Brownian Motion & More Wide & Other & 0\\

Brownian Motion & More Wide & Blank & 0\\

Brownian Motion & More Wide & Limitations & 3\\

Brownian Motion & More Wide & Principles & 5\\

Brownian Motion & More Wide & Principles + Limitations & 0\\

Brownian Motion & Same & Statistics & 14\\

Brownian Motion & Same & Limitations & 10\\

Brownian Motion & Same & Other & 1\\

Brownian Motion & Same & Principles & 9\\

Brownian Motion & Same & Blank & 6\\

Brownian Motion & Same & Principles + Limitations & 0\\

Brownian Motion & Single Value & Statistics & 0\\

Brownian Motion & Single Value & Limitations & 0\\

Brownian Motion & Single Value & Other & 0\\

Brownian Motion & Single Value & Principles & 0\\

Brownian Motion & Single Value & Blank & 0\\

Brownian Motion & Single Value & Principles + Limitations & 0\\
\hline
Single- Slit & More Narrow & Statistics & 8\\

Single- Slit & More Narrow & Principles + Limitations & 0\\

Single- Slit & More Narrow & Limitations & 2\\

Single- Slit & More Narrow & Other & 0\\

Single- Slit & More Narrow & Principles & 1\\

Single- Slit & More Narrow & Blank & 5\\

Single- Slit & More Wide & Statistics & 2\\

Single- Slit & More Wide & Principles + Limitations & 0\\

Single- Slit & More Wide & Limitations & 2\\

Single- Slit & More Wide & Other & 0\\

Single- Slit & More Wide & Principles & 1\\

Single- Slit & More Wide & Blank & 0\\

Single- Slit & Same & Principles + Limitations & 0\\

Single- Slit & Same & Statistics & 15\\

Single- Slit & Same & Other & 0\\

Single- Slit & Same & Blank & 6\\

Single- Slit & Same & Limitations & 12\\

Single- Slit & Same & Principles & 5\\

Single- Slit & Single Value & Principles + Limitations & 0\\

Single- Slit & Single Value & Statistics & 0\\

Single- Slit & Single Value & Other & 0\\

Single- Slit & Single Value & Blank & 1\\

Single- Slit & Single Value & Limitations & 0\\

Single- Slit & Single Value & Principles & 0\\
\hline\hline
\end{tabular}
\end{table*}

\begin{table*}
\caption{Raw frequency data associated with Fig.~\ref{fig: bbe}.}\label{ta:raw3}
\begin{tabular}{lllr}
\hline\hline
Scenario & Response & Code & Freq\\
\hline
Projectile Motion & More Narrow & Blank & 10\\

Projectile Motion & More Narrow & Statistics & 3\\

Projectile Motion & More Narrow & Principles & 1\\

Projectile Motion & More Narrow & Principles + Limitations & 3\\

Projectile Motion & More Narrow & Limitations & 81\\

Projectile Motion & More Narrow & Other & 16\\

Projectile Motion & More Wide & Blank & 0\\

Projectile Motion & More Wide & Statistics & 0\\

Projectile Motion & More Wide & Principles & 0\\

Projectile Motion & More Wide & Principles + Limitations & 0\\

Projectile Motion & More Wide & Limitations & 0\\

Projectile Motion & More Wide & Other & 0\\

Projectile Motion & Same & Principles & 0\\

Projectile Motion & Same & Blank & 6\\

Projectile Motion & Same & Principles + Limitations & 0\\

Projectile Motion & Same & Statistics & 1\\

Projectile Motion & Same & Other & 3\\

Projectile Motion & Same & Limitations & 4\\

Projectile Motion & Single Value & Principles & 0\\

Projectile Motion & Single Value & Blank & 2\\

Projectile Motion & Single Value & Principles + Limitations & 0\\

Projectile Motion & Single Value & Statistics & 1\\

Projectile Motion & Single Value & Other & 2\\

Projectile Motion & Single Value & Limitations & 17\\
\hline
Stern- Gerlach & More Narrow & Other & 0\\

Stern- Gerlach & More Narrow & Principles & 0\\

Stern- Gerlach & More Narrow & Blank & 6\\

Stern- Gerlach & More Narrow & Principles + Limitations & 3\\

Stern- Gerlach & More Narrow & Statistics & 1\\

Stern- Gerlach & More Narrow & Limitations & 21\\

Stern- Gerlach & More Wide & Other & 1\\

Stern- Gerlach & More Wide & Principles & 2\\

Stern- Gerlach & More Wide & Blank & 1\\

Stern- Gerlach & More Wide & Principles + Limitations & 0\\

Stern- Gerlach & More Wide & Statistics & 1\\

Stern- Gerlach & More Wide & Limitations & 1\\

Stern- Gerlach & Same & Other & 0\\

Stern- Gerlach & Same & Blank & 4\\

Stern- Gerlach & Same & Statistics & 0\\

Stern- Gerlach & Same & Principles & 9\\

Stern- Gerlach & Same & Principles + Limitations & 0\\

Stern- Gerlach & Same & Limitations & 6\\

Stern- Gerlach & Single Value & Other & 1\\

Stern- Gerlach & Single Value & Blank & 0\\

Stern- Gerlach & Single Value & Statistics & 0\\

Stern- Gerlach & Single Value & Principles & 0\\

Stern- Gerlach & Single Value & Principles + Limitations & 0\\

Stern- Gerlach & Single Value & Limitations & 3\\
\hline\hline
\end{tabular}
\quad
\begin{tabular}{lllr}
\hline\hline
Scenario & Response & Code & Freq\\
\hline
Brownian Motion & More Narrow & Statistics & 3\\

Brownian Motion & More Narrow & Other & 1\\

Brownian Motion & More Narrow & Blank & 4\\

Brownian Motion & More Narrow & Limitations & 16\\

Brownian Motion & More Narrow & Principles & 1\\

Brownian Motion & More Narrow & Principles + Limitations & 7\\

Brownian Motion & More Wide & Statistics & 0\\

Brownian Motion & More Wide & Other & 0\\

Brownian Motion & More Wide & Blank & 1\\

Brownian Motion & More Wide & Limitations & 0\\

Brownian Motion & More Wide & Principles & 1\\

Brownian Motion & More Wide & Principles + Limitations & 0\\

Brownian Motion & Same & Statistics & 2\\

Brownian Motion & Same & Limitations & 1\\

Brownian Motion & Same & Other & 0\\

Brownian Motion & Same & Principles & 13\\

Brownian Motion & Same & Blank & 6\\

Brownian Motion & Same & Principles + Limitations & 3\\

Brownian Motion & Single Value & Statistics & 1\\

Brownian Motion & Single Value & Limitations & 0\\

Brownian Motion & Single Value & Other & 0\\

Brownian Motion & Single Value & Principles & 0\\

Brownian Motion & Single Value & Blank & 0\\

Brownian Motion & Single Value & Principles + Limitations & 0\\
\hline
Single- Slit & More Narrow & Statistics & 3\\

Single- Slit & More Narrow & Principles + Limitations & 3\\

Single- Slit & More Narrow & Limitations & 13\\

Single- Slit & More Narrow & Other & 2\\

Single- Slit & More Narrow & Principles & 3\\

Single- Slit & More Narrow & Blank & 5\\

Single- Slit & More Wide & Statistics & 0\\

Single- Slit & More Wide & Principles + Limitations & 0\\

Single- Slit & More Wide & Limitations & 0\\

Single- Slit & More Wide & Other & 0\\

Single- Slit & More Wide & Principles & 1\\

Single- Slit & More Wide & Blank & 1\\

Single- Slit & Same & Principles + Limitations & 3\\

Single- Slit & Same & Statistics & 3\\

Single- Slit & Same & Other & 0\\

Single- Slit & Same & Blank & 6\\

Single- Slit & Same & Limitations & 3\\

Single- Slit & Same & Principles & 11\\

Single- Slit & Single Value & Principles + Limitations & 0\\

Single- Slit & Single Value & Statistics & 0\\

Single- Slit & Single Value & Other & 0\\

Single- Slit & Single Value & Blank & 2\\

Single- Slit & Single Value & Limitations & 1\\

Single- Slit & Single Value & Principles & 0\\
\hline\hline
\end{tabular}

\end{table*}

\end{document}